\documentclass[twocolumn,nofootinbib,showpacs,prd]{revtex4}

\usepackage{graphics,graphicx}
\usepackage{amsmath}
\usepackage{psfrag}
\usepackage{dcolumn}
\usepackage{color}

\def\no{\nonumber \\}

\newcommand{\be}{\begin{equation}}
\newcommand{\ee}{\end{equation}}
\newcommand{\bea}{\begin{eqnarray}}
\newcommand{\eea}{\end{eqnarray}}
\def\no{\nonumber \\}
\newcommand{\mc}{\mathcal}
\newcommand{\vek}[1]{\boldsymbol{#1}}

\def\hxqt{      h_{\times}|_Q(t)        }

\begin{document}

\title{
Gravitational waves from compact binaries inspiralling along post-Newtonian accurate
eccentric orbits: Data analysis implications
}

\author{Manuel Tessmer}
\email{M.Tessmer@uni-jena.de}
\affiliation{Theoretisch-Physikalisches Institut,
Friedrich-Schiller-Universit\"at Jena, 
Max-Wien-Platz 1,
07743 Jena, Germany}

\author{Achamveedu Gopakumar}
\email{A.Gopakumar@uni-jena.de}
\affiliation{Theoretisch-Physikalisches Institut,
Friedrich-Schiller-Universit\"at Jena, 
Max-Wien-Platz 1,
07743 Jena, Germany}

\date{\today}

\begin{abstract}
Compact binaries inspiralling along eccentric orbits are plausible
gravitational wave (GW)
sources  for the ground-based laser interferometers. We explore the losses in the event rates incurred
when searching for GWs from compact binaries inspiralling along post-Newtonian accurate
eccentric orbits with certain obvious non-optimal search templates.
For the present analysis, GW signals having 2.5 post-Newtonian accurate
orbital evolution
are modeled following the phasing formalism,
presented in
[T.~Damour, A.~Gopakumar, and B.~R.~Iyer,
Phys. Rev. D \textbf{70}, 064028 (2004)]. 
We demonstrate that the search templates that model in a gauge-invariant manner GWs
from compact binaries inspiralling under qudrupolar radiation reaction along
2PN accurate circular orbits are very efficient in capturing
our somewhat realistic GW signals.
However, three types of search templates based on the adiabatic, complete adiabatic
and gauge-dependent complete non-adiabatic approximants, detailed in
[P.~Ajith, B.~R.~Iyer, C.~A.~K.~Robinson and B.~S.~Sathyaprakash,
Phys.\ Rev.\  D {\bf 71}, 044029 (2005)],
relevant for the circular inspiral under the qudrupolar radiation reaction
were found to be inefficient
in capturing the above mentioned eccentric signal.
We conclude that further investigations 
will be required to probe the ability of 
various types of PN accurate circular templates, employed to analyze the LIGO/VIRGO data,
to capture GWs from compact binaries having tiny orbital eccentricities. 

\end{abstract}

\pacs{
04.30.Db, 
04.25.Nx 
}

\maketitle

\section{Introduction}

Compact binaries, namely neutron star--neutron star, black hole--neutron star and
black hole--black hole binaries,
inspiralling in quasi-circular orbits are the most plausible sources of 
gravitational radiation for the first generation 
ground-based laser interferometric 
gravitational-wave (GW)
detectors ~\cite{Abbott:2007wu,Hild:2006bk,Acernese2006}.
GW data analysis
communities, analyzing noisy data from the operating interferometers,  require accurate and
efficient temporally evolving GW 
polarizations,
$h_{+} (t)$ and $h_{\times}(t)$, the so called GW search templates.  In the ongoing efforts to construct
GW templates, inspiralling compact binaries 
are modeled as point particles moving in quasi-circular orbits \cite{BDI}.
The approximation of quasi-circularity in the orbital description while constructing GW templates
is quite appropriate, because gravitational radiation reaction
quickly reduces the orbital eccentricity as a compact binary evolves
towards its last stable orbit
(LSO).  Employing the dominant contributions to energy and angular
momentum losses via GW emission for an inspiralling compact binary,
it is straightforward to deduce that
when its semi major axis is halved, its eccentricity roughly reduces by a factor of three \cite{P64}.
The above argument implies that the orbital eccentricity of the Hulse-Taylor binary 
pulsar when its orbital frequency reaches around $20$ Hz will be $\sim 10^{-6}$. 
Further, it was argued that the templates, constructed to detect GWs from 
inspiralling compact binaries in quasi-circular orbits
should be quite successful in extracting GWs from (mildly) eccentric binaries \cite{MP00}.
Therefore, at present various GW data analysis communities are not explicitly searching for GWs from 
compact binaries in inspiralling eccentric orbits.

In this paper, we revisit the issue addressed in Ref.~\cite{MP00}, namely the
possibility of rather efficiently extracting GWs from compact binaries in
inspiralling eccentric orbits using search templates constructed for compact
binaries evolving in quasi-circular orbits.
In our opinion, it is justified to doubt the conclusions of Ref.~\cite{MP00} 
because it is
not that difficult to note that GWs from compact
binaries in non-circular orbits were not accurately modeled in Ref.~\cite {MP00},
even while restricting the radiation reaction to the dominant order.
We observed that dominant secular, but non-reactive, contributions to the orbital 
evolution were
ignored in Ref.~\cite{MP00} (see our Section~\ref{sec:MP00} for details). Moreover, 
several astrophysically motivated
investigations, carried out in the last few years, indicate that 
compact binaries in inspiralling eccentric orbits
are plausible sources of GWs even for the ground-based GW detectors \cite{e_scenarios}.
Therefore, it is rather important to explore the ability of various types of 
circular templates to capture somewhat  realistic modeling of GWs from 
compact binaries in inspiralling eccentric orbits.    
However, we would like to point out that the 
ability of various types of circular  inspiral search templates,
available in the LSC Algorithms Library (LAL) \cite{LAL},
to capture GW signals from inspiralling compact binaries having tiny orbital eccentricities
will require further investigations.
This is due to the fact that in this paper, and following Ref.~\cite{MP00},
we only include the dominant contributions to the reactive dynamics.

       This paper is organized in the following way. In  Section~\ref{sec:MP00}, 
we explicate the reasons for our current 
investigation. Section~\ref{sec: 2.5_GW_phasing}
presents a brief summary of constructing accurate gravitational waveforms
for compact binaries having arbitrary mass ratio moving in inspiralling eccentric orbits. 
How we revisit the analysis,
presented in Ref.~\cite{MP00}, is detailed in  Section~\ref{sec: NO_matched_filter}
and  we also present our results in this section. 
A brief summary, our conclusions and future directions are provided in Section~\ref{sec: conclusion}.
Appendix~\ref{phasing:Appendix:A}
deals with certain computational details.


\section{Residual orbital eccentricity and its implications }
\label{sec:MP00}
The GW signals, namely $h_{+}(t)$ and $h_{\times}(t)$,
emitted by inspiralling compact binaries can be accurately computed 
employing the post-Newtonian (PN)
approximation to general relativity. 
The PN approximation to the dynamics of inspiralling compact binaries,
usually modeled to consist of point masses, 
provides, for example,
the equations of motion  as corrections to the Newtonian one
in terms of
 $({v}/{c})^2 \sim {G m}/{c^2\,r}$, 
where $v$, $m$ and $r$ are
the characteristic orbital velocity,
the total mass, and the typical orbital separation,
respectively. 
The way
any residual orbital  eccentricity influences possible loss of event rate for the initial LIGO
was explored, for the first time, in Ref.~\cite{MP00}.
This was achieved by computing the drops in the signal-to-noise-ratio 
while searching for GWs from compact binaries moving in inspiralling
eccentric orbits with templates constructed for binaries in quasi-circular orbits.
The search templates employed in Ref.~\cite{MP00} are essentially given by the following
expressions:
\begin{align}
\label{hx_circ}
h_{\times} ( \phi, \omega) |_R = & -4 C \frac{(G m \eta) }{c^2 R'} \biggl ( 	\frac{G m \omega}{c^3}	\biggr )^{2/3}
 \sin 2\phi \,,
\end{align}
where the familiar symbols $m $ and 
$\eta$ stand for the total mass and  the symmetric mass ratio, while $C $ and $R'$ denote 
the cosine of the orbital inclination and the radial distance to the binary.
The subscript $R$ appearing on the left hand side of Eq.~(\ref{hx_circ}) indicates that
we are using the so-called restricted PN waveforms as the search templates.
The temporal evolution for $\phi$ and $\omega$, the orbital phase and angular frequency,
are governed by
\begin{subequations}
\label{omega_evol}
\begin{align}
\label{Eq.2a}
\frac{d \phi}{dt} &= \omega\,,   \\
\label{Eq.2b}
\frac{d \omega}{dt} & \equiv
\frac{{\cal L(\omega)}_N}{ d {\cal E}_N/d \omega} =
 \frac{96}{5}
\left ( \frac{ G\, {\cal M}\,\omega}{c^3} \right )^{5/3}
\omega^2 \,,
\end{align}
\end{subequations}
where ${\cal L(\omega)}_N$ and ${\cal E}_N$ stand for the dominant contributions to the GW luminosity
and the Newtonian orbital energy and the chirp mass ${\cal M} \equiv m\,\eta^{3/5}$.
Further,
it is customary to use $ \omega = \pi\, f_{\rm GW}$, $ f_{\rm GW}$ being the
GW frequency to provide the limits for $\omega$.
In the GW literature, Eqs.~(\ref{hx_circ})  and (\ref{omega_evol}) provide
the `Newtonian templates' for the quasi-circular inspiral
due to the fact that only the dominant contributions to ${\cal L(\omega)}$ and ${\cal E}(\omega)$ are
required to construct $d \omega/dt$.
However, it is interesting to note that Eqs.~(\ref{omega_evol}) provide
certain 2.5PN accurate orbital phase evolution.
This observation is due to the fact that in Eqs.~(\ref{omega_evol}), one perturbes
a compact binary in an exact circular orbit, defined by Eq.~(\ref{Eq.2a}),
by an expression for $d \omega/dt$, given by Eq.~(\ref{Eq.2b}).
Therefore, it is reasonable to state that
Eqs.~(\ref{hx_circ})  and (\ref{omega_evol}) provide a
prescription to obtain 2.5PN accurate
GW phasing for compact binaries inspiralling along exact circular orbits.
We observe that Ref.~\cite{MP00}
employed analytically given Fourier domain version of the above search templates
(see Eqs.~(13), (14) and (15) in Ref.~\cite{MP00}).

  Following Ref.~\cite{AIRS}, there exists two more ways of constructing
circular inspiral templates that incorporate the Newtonian 
reactive dynamics and we detail them below.
The orbital phase evolution under the so-called complete adiabatic approximant
at the Newtonian order radiation reaction reads
\begin{subequations}
\label{Eq.3}
\begin{align}
\label{Eq.3a}
\frac{d \phi}{dt} &= \omega\, \equiv \frac{c^3}{G\,m}\, x^{3/2}\,,   \\
\label{Eq.3b}
\frac{d x}{dt} & \equiv
\frac{{\cal L}(x)_N}{ d {\cal E}(x)_{2PN}/d x} \,, \mbox{where}
\\
\label{Eq.3c}
{\cal L}(x)_N &=
\frac{32\,\eta^2\,c^5}{5\,G}\, x^{5}
\,,\\
\label{Eq.3d}
{\cal E}(x)_{2PN} &=
 -\frac{\eta\, m\, c^2}{2}\,x
\biggl \{
1
- \frac{1}{12} \biggl [ 9 +  \eta \biggr ] x
\no
&
+
\biggl [
-{\frac {27}{8}}
+{ \frac {19}{8}}\,\eta
-\frac{1}{24}\,{\eta}^{2}
\biggr ]{x}^{2}
\biggr \}
\,.
\end{align}
\end{subequations}
The crucial difference between Eq.~(\ref{omega_evol}) and (\ref{Eq.3}) is that
the numerator in Eq.~(\ref{Eq.3b}) is 2PN accurate and the use of 
the $x$ variable: $x \equiv (G\,m\, \omega/c^3)^{2/3}$.
Following Ref.~\cite{AIRS}, we do not expand the right hand side of Eq.~(\ref{Eq.3b})
and therefore, Eqs.~(\ref{Eq.3}) also provide a prescription to compute
2.5PN order orbital phase evolution.
We recall that in the language of Ref.~\cite{AIRS}, Eq.~(\ref{omega_evol})
provide the adiabatic approximant
at the Newtonian order.

 The next prescription provides the complete non-adiabatic approximant at the 2.5PN order and
following \cite{AIRS}, its orbital evolution is defined by
\begin{subequations}
\label{Eq.4}
\begin{align}
\label{Eq.4a}
\frac{d \phi}{dt} &= \omega = \frac{ v}{r}\,,   \\
\label{Eq.4b}
\frac{d \vek {y} }{dt} & \equiv {\vek v }\,,
\\
\label{Eq.4c}
\frac{d \vek {v} }{dt} &=
-\frac{G\,m}{r^3}\, \biggl \{ 1 - \biggl [ 3 -\eta \biggr ] \,\frac{G\,m}{c^2\,r} 
+ \biggl [ 6 + \frac{41}{4}\,\eta 
\no
&
+ \eta^2 \biggr ]\, (\frac{G\,m}{c^2\,r})^2 \biggr \} \, {\vek y }
-\frac{32}{5}\, \eta \, \frac{G^3\, m^3}{c^5\, r^4}\, {\vek v }
\,,
\end{align}
\end{subequations}
where $\vek y $ and $\vek v $ define the orbital separation and velocity vectors.
It is important to note that the complete non-adiabatic approximant is gauge-dependent
and we employed the harmonic gauge in Eqs.~(\ref{Eq.4}).
Further, note that we are using the circular limit of Eqs.~(3.1) to (3.6) in Ref.~(\cite{AIRS})
as we are interested in constructing only circular templates.

 Let us now take a closer look at the way 2.5PN accurate GW phasing for 
eccentric binaries, required to model
GW signals from compact binaries in inspiralling eccentric orbits,
were performed in Ref.~\cite{MP00}.
%
%
%
In what follows, first we describe a fully PN accurate prescription to 
construct the above mentioned GW signals.
The process of constructing the expected, and therefore PN accurate, 
GW signals from compact binaries inspiralling along eccentric orbits,
detailed in Ref.~\cite{DGI}, is rather involved. 
To begin with, GW phasing requires an expression, similar to Eq.~(\ref{hx_circ}), for  $h_{\times}$.
The dominant quadrupolar contribution to $h_{\times}$, denoted by $h_{\times}\big|_{\rm Q}$,
reads
\begin{align}
\label{hx_ecc}
h_{\times}(r,\phi,\dot{r},\dot{\phi}) \big|_{\rm Q}
&=-\frac{2 G m \eta C}{c^4 R'}
\bigg[
\bigg( \frac{G m}{r} + r^2 \dot{\phi}^2 - \dot{r}^2 \bigg) \sin 2 \phi
\nonumber
\\
& \quad
- 2 \dot{r} r \dot{\phi} \cos 2 \phi
\bigg]
\,,
\end{align}
where 
$\dot r = dr/dt$  and $ \dot \phi = d \phi/dt $.
When one is interested in doing 2.5PN accurate GW phasing for eccentric binaries, 
following Ref.~\cite{DGI},  
it can be argued that the
dominant secular reactive evolutions for the dynamical variables, 
 $r$, $\phi$, $\dot r$ and $\dot \phi$,
are governed by the
following two differential equations:
\begin{subequations}
\label{n_et_evol_correct}
\begin{align}
\frac{d {n}}{dt} & =
\frac{ \xi^{5/3} n^2 \eta }{ 5 (1 - e_t^2)^{7/2} }
\Big \{ 96 + 292 e_t^2 + 37 e_t^4 \Big \}
\,,
\\
\frac{d {e}_t}{dt} & =
- \frac{ \xi^{5/3} n \eta e_t }{ 15 (1 - e_t^2)^{5/2} }
\Big \{ 304 + 121 e_t^2 \Big \}
\,,
\end{align}
\end{subequations}
where $n =2\,\pi/ T_r $ and $e_t$ are the so-called mean motion and time eccentricity (and
$T_r$ being the radial orbital period) and $\xi$ is a shorthand notation for $G m n/c^3$.
These orbital elements naturally arise 
in the PN accurate Keplerian type parametric solution to the conservative PN accurate compact
binary dynamics, available in Refs.~\cite{DD,DS88,WS,MGS}.
However, the orbital phase of an eccentric binary secularly evolves partly due to the advance of
periastron and this effect appears at the first and second PN (conservative) orders.
Therefore, in order to have 2.5PN accurate orbital phase evolution, it
is imperative to include secular non-radiative effects appearing at the first and  second
PN orders. In Ref.~\cite{DGI}, this is achieved by having the following 2PN accurate
expressions for $\phi$, symbolically written as
\begin{subequations}
\label{phi_split}
\begin{align}
\phi &= \lambda + W (l;n,e_t)\,,\\
\lambda &= ( 1 + k ) n ( t- t_0) + c_{\lambda}\,,
\\ \mbox{with} \,\,\,\, 
l &=n ( t- t_0) + c_l \,.
\end{align}
\end{subequations}
In above equations $k = \frac{ \Delta \Phi}{2\, \pi} $, $\Delta \Phi$  being the advance of periastron
in the time interval $T_r$. 
Following Ref.~\cite{DGI}, we note that $ W (l;n,e_t)$ is a 2PN accurate function, given explicitly
in terms of eccentric anomaly $u$, $n$ and $e_t$ and the mean anomaly $l $
is related to $u$ by the 2PN accurate Kepler Equation (KE).
The constants $t_0$, $c_l$ and $ c_{\lambda}$ refer to some initial instant and values
of $l$ and ${\lambda}$ at $t= t_0$ (the explicit 2PN accurate expressions for
$\lambda (l; n, e_t) $  and $ W (l;n,e_t)$ in harmonic gauge is available in Eqs.~(25) in
Ref.~\cite{KG06}).
The reactive secular changes in $\phi(t)$ enter via $n(t)$ and $e_t(t)$ and these time
dependencies are governed by Eqs.~(\ref{n_et_evol_correct}). To be consistent in a PN way,
it is also desirable to employ 2PN accurate parameteric equations for $r, \dot r, \phi $ and
$\dot \phi$ in terms of $n, e_t$ and $u(l;n,e_t)$ (see Ref.~\cite{DGI} for more details).
Therefore to investigate if search templates constructed for compact binaries in
quasi-circular orbits, given by Eqs.~(\ref{hx_circ}), (\ref{omega_evol}),
(\ref{Eq.3}) and (\ref{Eq.4})
are good enough to detect, via matched filtering, 
GWs from binaries in inspiralling eccentric orbits, 
it is desirable to model GW signals following Ref.~\cite{DGI}.
In other words, GW signals should be constructed 
incorporating 2.5PN accurate orbital motion in Eq.~(\ref{hx_ecc}).

\begin{figure}[ht]
\includegraphics[height=8.5cm, width=5.cm, angle=-90]{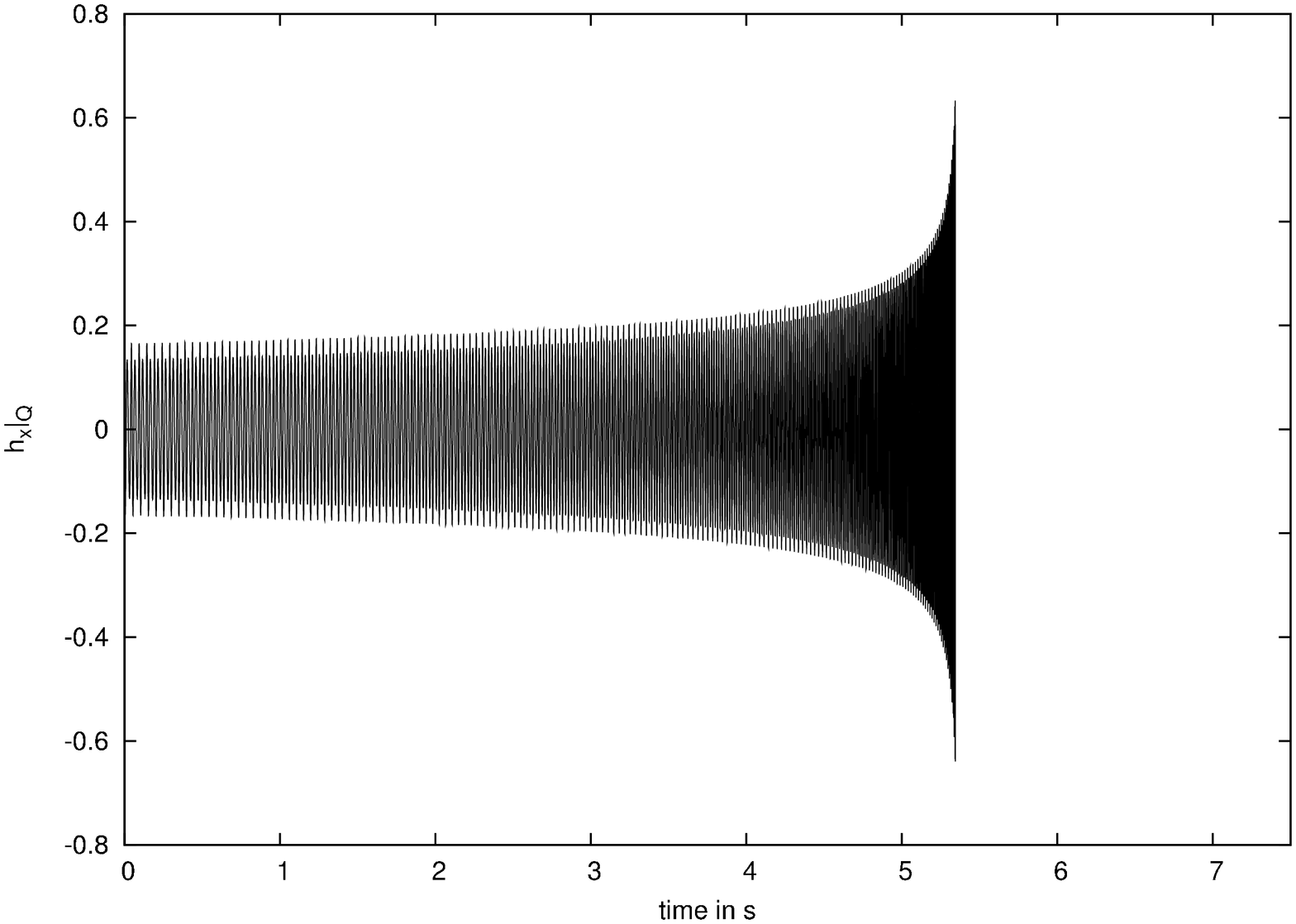}\\[0.1cm]
\includegraphics[height=8.5cm, width=5.cm, angle=-90]{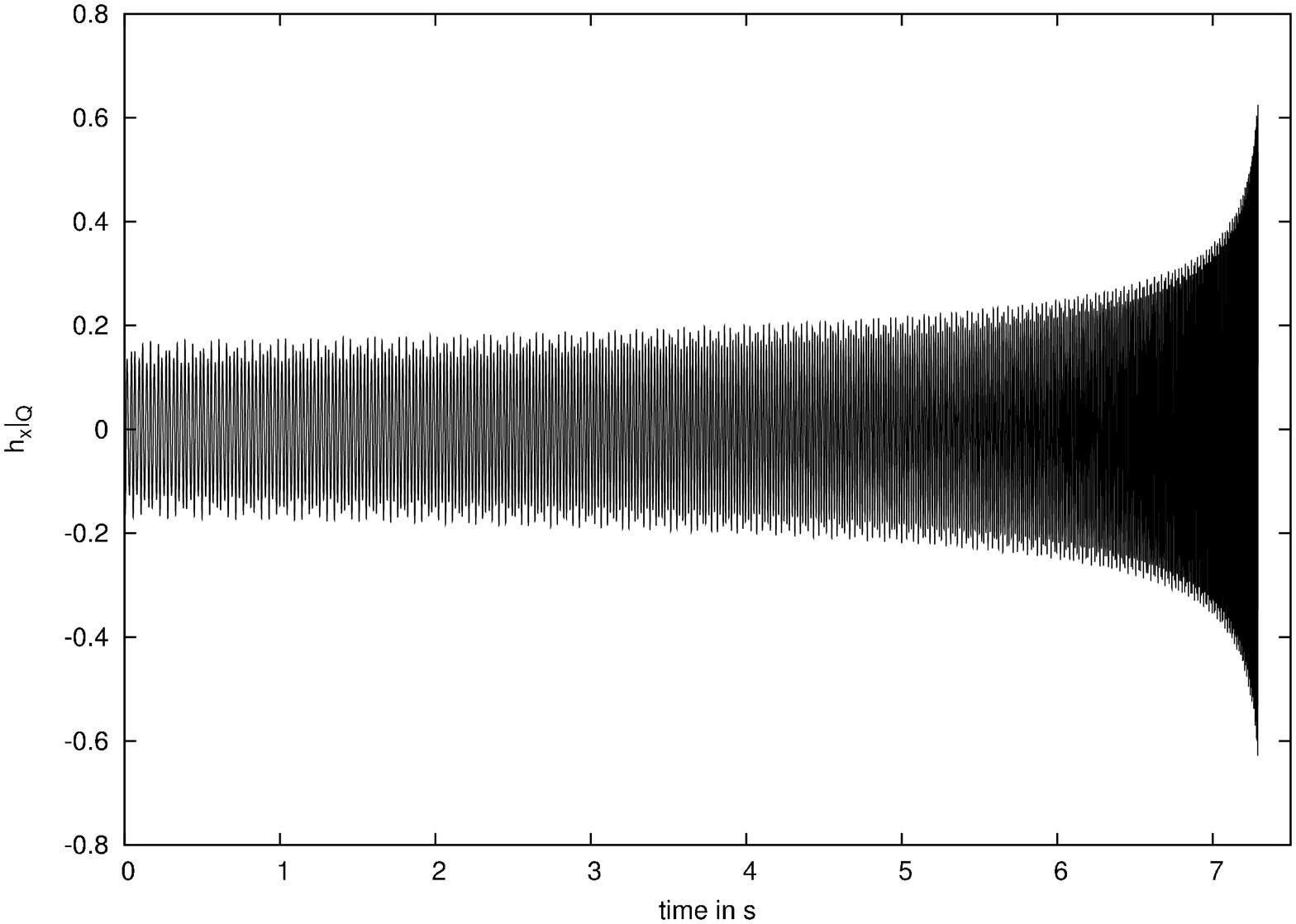}
\caption{\label{fig:different_evol} 
Plots showing temporal evolution of scaled $h_{\times}|_Q (t)$, having $t$ in
seconds, for $m_1 = 1.4 M_{\odot}, m_2 = 10 M_{\odot} $ and initial $e_t = 0.1$.
For the upper plot, orbital motion is governed by Eqs.~(\ref{n_et_evol_correct})
and (\ref{orbital_elmts_NEWT}) and we have $n_i=125.7$Hz and $n_f=1211.8$Hz.
For the lower plot, the orbital motion is 2.5PN accurate and $n_i= 111.7$Hz and $n_f=741.0$Hz,
obtainable with the help of Eq.~(\ref{omega_n_2PN}).
The two sets of initial and final $n$ values result from the fact that for both
cases quadrupolar $f_{GW}$ can only vary between $40$Hz and $( 6^{3/2}\, \pi\, G\, m/c^3 )^{-1}$.
Further, notice the visible effect of $k$ in the lower plot.
}
\end{figure}

  However, this is not the way GWs from inspiralling eccentric orbits are modeled in
Ref.~\cite{MP00} and one can summarize their inaccurate construction of GW signals 
in the following way.
We observe that
Ref.~\cite{MP00} essentially employs the same expression for $h_{\times}|_{\rm Q}$, but
descriptions for conservative $r, \dot r, \phi $ and $\dot \phi$ were purely Newtonian and
radiation reaction enters via Eqs.~(\ref{omega_evol}) (we recall that Ref.~\cite{MP00}
employs different orbital variables to describe the binary dynamics and therefore required to
solve numerically three coupled differential equations). The crucial difference
with respect to what is done in Ref.~\cite{DGI}
is that the secular 1PN and 2PN corrections to the orbital phase
evolution
were neglected in Ref.~\cite{MP00}. 
The conservative orbital dynamics pursued in Ref.~\cite{MP00} may be defined in terms 
of Keplerian parameterization and it reads
\begin{subequations}
\label{orbital_elmts_NEWT}
\begin{align}
r	  =&	\left( \frac{G m}{n^2} \right)^{1/3} ( 1 - e_t \cos u )		\,, \\
\dot{r}   =&	\frac{ e_t (G m n)^{1/3} }{(1 - e_t \cos u)}\, \,\sin u		\,, \\
\label{phi_NEWT}
\phi	  =&	\lambda + W(l) = n(t-t_0) + (v-u) + e_t \sin u			\,, \\
\dot \phi =& \frac{ n \sqrt{1 - e_t^2} }{ (1 - e_t \cos u )^2 }			\,,
\\ \mbox{where} \,\,\,\, 
v - u  =& 2 \tan^{-1}
\left(
\frac{ \beta \sin u }{ 1 - \beta \cos u }
\right)
\,,
\end{align}
\end{subequations}
and  $\beta $ being  $( 1 - \sqrt{ 1 - e_t^2 } ) / e_t$. In above expressions, 
$u, v$ and $e_t$  are the usual eccentric and true anomalies and Newtonian orbital eccentricity
of the Keplerian parameterization.
The reactive evolution (again) is governed by Eqs.~(\ref{n_et_evol_correct}) and
the explicit temporal evolution of $h_{\times}|_Q (t)$  also
requires numerical solution of classical Kepler Equation $ l \equiv n(t - t_0) = u - e_t\, \sin u$.
The above mentioned two approaches to obtain $h_{\times}|_Q (t)$ lead to quite different GW signals for
compact binaries moving in inspiralling eccentric orbits as demonstrated in Fig.~\ref{fig:different_evol}.
For $h_{\times}|_Q (t)$, constructed following Ref.~\cite{DGI}, 
one can clearly observe
secularly changing modulations due to periastron advance.
Another feature to note is that $h_{\times}|_Q (t)$, having 2.5PN accurate orbital motion,
lasts longer than $h_{\times}|_Q (t)$ having Eqs.~(\ref{n_et_evol_correct}) imposed on
Newtonian accurate orbital motion.
This is mainly because of using different $n_i$ and $n_f$, initial and final values for $n$,
for the two cases. This is 
required 
to make sure that the initial and final frequencies of the
emitted quadrupolar GWs are 40 Hz and $ ( 6^{3/2} \, \pi\, G m/c^3 )^{-1}$ for our two cases
(recall that advance of periastron shifts the dominant harmonic as explained in Ref.~\cite{TG06}).
Further, while constructing these figures, we assume that first and third harmonics are 
irrelevant for the initial LIGO.
In our opinion, astrophysical GWs from compact binaries in inspiralling eccentric orbits 
are rather more accurately (and realistically) modeled in Ref.~\cite{DGI} compared to Ref.~\cite{MP00}
and therefore, we are justified to doubt conclusions presented in Ref.~\cite{MP00}.
However, it is important to note that in a purely
GW data analysis sense, while employing Eqs.~(\ref{hx_ecc}), (\ref{n_et_evol_correct})
and (\ref{orbital_elmts_NEWT}) to model GWs from
inspiralling eccentric binaries,
the conclusions of  Ref.~\cite{MP00} are indeed correct. This allowed us to
accurately calibrate our codes with the results presented in Ref.~\cite{MP00}.

   We are now in a position to explore if the various circular templates,
given by Eqs.~(\ref{hx_circ}), (\ref{omega_evol}), (\ref{Eq.3}) and (\ref{Eq.4}),
are going to efficient in capturing GWs from eccentric binaries.
Because the
matched filtering requires accurate
secular description for GW phase evolution, this can roughly be done by computing and
comparing (for the various cases under study)
$\mathcal N_{GW}$, the number of accumulated
GW cycles between some minimum and maximum GW frequencies.
Making use of the definition $\mathcal N_{GW} = ( \phi_{\rm max} - \phi_{\rm min})/{\pi}$,
where $\phi_{\rm max}$ and $\phi_{\rm min}$ are the values of the orbital phase
associated with the some minimum and maximum GW frequencies, we can easily evaluate
the $\mathcal N_{GW}$ associated with an approximant in the time-domain.
We list below the various cases involved in the present study,
where the reactive effects are restricted to the dominant quadrupolar contributions.
\begin{itemize}
\item
Case~I: The adiabatic approximant and the
associated search templates are
given by Eqs.~(\ref{hx_circ}) and ~(\ref{omega_evol}).
\item
Case~II: The complete adiabatic circular inspiral and the associated
search templates are specified by Eqs.~(\ref{hx_circ}) and (\ref{Eq.3}).
\item
Case~III: The gauge-dependent complete non-adiabatic circular inspiral and the associated
search templates are specified by Eqs.~(\ref{hx_circ}) and (\ref{Eq.4}).
\item
Case~IV: The eccentric $h_{\times}(t)$, given by Eqs.~ (\ref{hx_ecc}),
(\ref{orbital_elmts_NEWT}) and (\ref{n_et_evol_correct}) and this is
the GW signal employed in Ref.~\cite{MP00}.
\item
Case~V: The eccentric $h_{\times}(t)$, given by Eqs.~ (\ref{hx_ecc}) and where
the 2.5PN accurate orbital phase evolution is computed following Ref.~\cite{DGI}.
\item
Case~VI: The newly introduced circular templates given by 
Eqs.~~(\ref{quasi_c_hx}) and (\ref{quasi_c_dt_phi}),
\end{itemize}
 
 We would like to point out that Ref.~\cite{MP00} investigated a scenario 
involving the cases IV and I. In this study, we use the case V to model the 
fiducial GWs from an eccentric binary and employ the cases I, II, III and VI
to construct the inspiral templates[recall that cases II, III and VI do 
incorporate in different ways 2PN accurate conservative dynamics 
to construct the search templates]. 
The use of the case I to model search templates 
in our comparison is based on the following two arguments. The first is that,
strictly speaking and as explained earlier, Eqs.~(\ref{omega_evol}) also
describe 2.5PN accurate orbital phase evolution of
a compact binary inspiralling along
exact circular orbits, defined by Eq.~(\ref{Eq.2a}). 
The second reason is that we would like to probe the performances of all the three types 
of search templates, introduced in Ref.~\cite{AIRS} and having quadrupolar reactive dynamics, 
while treating the case V to model GWs from eccentric binaries.  

  Let us briefly explain how we perform the 
$\mathcal N_{GW}$ computations in the time-domain
for the various cases. For the case I, we use $40 \, \pi$Hz and 
$ \left( 6^{3/2}\, G m/c^3 \right )^{-1}$ as the 
initial and final values of $\omega$ for the temporal evolution defined by Eqs.~(\ref{omega_evol})
and similar limits also apply to the case II.
The difference in $\phi$ at the above two $\omega$ values leads to $\mathcal N_{GW}$ for 
the cases I and II.
For the case III, we need to specify $r$ and $v$ values associated with the above mentioned 
$\omega$ limits and we use 2PN accurate expression for $\omega$ in terms of $r$, given by Eq.~(3.11) 
in Ref.~\cite{BDI95},
to obtain the relevant limits for $r$ and $v = \omega\,r$. 
For the Newtonian accurate conservative and eccentric orbital motion, treated as the case IV,
the lower and upper values of $n$, $n_i$ and $n_f$,
are identified with the above mentioned $\omega$ limits due to the fact that at this order
$n = \omega$. Further, reactive evolution is given by Eqs.~(\ref{n_et_evol_correct}) and the values of 
$\phi$, obtained by using relevant parts of Eqs.~(\ref{orbital_elmts_NEWT}) and the classical KE, evaluated 
at the instances when $n$ reaches the values $n_i$ and $n_f$ lead to $\mathcal N_{GW}$ for the case IV.
For the case V, the values for $n_i$ and $n_f$ are
numerically obtained by using the following 2PN accurate relation
\begin{align}
\label{omega_n_2PN}
\pi\, f_{\rm GW}	& = n	\biggl \{ 1+ \frac{ 3 \xi ^{2/3} }{ 1 - e_t^2 } 
 + \frac{ \xi ^{4/3} }{ 4 ( 1 - e_t^2 )^2 }
		\biggl [ 78  - 28 \eta 
\nonumber \\ &
+ ( 51 - 26 \eta ) e_t^2 \biggr ]
		\biggr \}\,.
\end{align}
This is justified because of the observation in Ref.~\cite{TG06} that the dominant GW spectral component
of a mildly eccentric compact binary, having PN accurate orbital motion, appears at
$( 1 + k) \, n/ \pi$.
Using $\omega_i = 40\,\pi$Hz and $ \omega_f = \left( 6^{3/2}\, G m/c^3 \right )^{-1 } $, it is easy to obtain numerically $n_i$ and $n_f$ for the 
three canonical binaries  considered in Table \ref{Table:1_NOC}.
Using 2.5 PN accurate $\phi(l;n, e_t)$ in harmonic gauge, 
obtainable using Eqs.~(25) and (27) in Ref.~\cite{KG06} and Eqs.~(\ref{n_et_evol_correct}),
we compute values of $\phi(t)$ at instances when $n$ reaches its above mentioned 
limiting values as $\phi_{\rm min}$ and
$\phi_{\rm max}$.   
It is not very difficult to infer that 
we can choose, without any loss of generality,
$\phi_{\rm min} =0$ at $t=0$ in all cases of $\mathcal N_{GW}$ evaluations.
Finally, for the case VI, the limiting values for $n$ are obtained using the circular limit of
Eq.~(\ref{omega_n_2PN}) and a procedure similar to the case I to compute the relevant
$\mathcal N_{GW}$.

     The resulting GW cycles, $\mathcal N_{GW}$, for three typical compact binaries
computed using the above mentioned six different prescriptions  are listed in
Table~\ref{Table:1_NOC}. We observe that the differences in $\mathcal N_{GW}$ between
cases I and IV are not very large. This indicates that quasi-circular templates, given by
Eqs.~(\ref{hx_circ}) and (\ref{omega_evol}), should be highly efficient to pick up GW
signals modeled using
Eqs.~(\ref{hx_ecc}), (\ref{n_et_evol_correct}) and (\ref{orbital_elmts_NEWT}) 
for small values of initial eccentricities.
We also observe that for a given $e_t$ the differences in
$\mathcal N_{GW}$ between the cases I and IV decrease as we increase the total mass. This is
also consistent with the observation in Ref.~\cite{MP00} that for high mass binaries
eccentricities up to 0.2 may be tolerated while searching with
quasi-circular templates.
However, the differences in $\mathcal N_{GW}$ between the cases I, II, III and V
are quite large even when initial $e_t \sim 0.01$.
This makes it interesting to revisit the analysis performed in Ref.~\cite{MP00}.
However, the differences in $\mathcal N_{GW}$ between the cases V and VI
indicate that the new type of circular inspiral templates, given by 
Eqs.~(\ref{quasi_c_hx}) and (\ref{quasi_c_dt_phi}), should be
more efficient in capturing our somewhat realistic eccentric GW signals.
In the next section, we briefly summarize 2.5PN accurate GW phasing, available in Ref.~\cite{DGI},
and present a prescription to perform it in a highly accurate and efficient manner.

\begin{table}[!ht]
\caption{
\label{Table:1_NOC}
The accumulated number of GW cycles, $\mathcal N_{GW}$,
relevant for the initial LIGO, while considering the above detailed six types 
of orbital phase evolutions for the three types of canonical binaries.
We recall that the radiation reaction is always restricted to the dominant
quadrupolar order.
}
\begin{tabular}{||l|r|r|r|}
\hline
 $m_1/ M_{\odot} : m_2/ M_{\odot} $ & $1.4 : 1.4$ & $1.4 : 10$ & $10 : 10$ \\
\hline
\multicolumn{4}{||c|}{ Case I		} \\
\hline
$e_t \equiv 0$ \hfill	   & 1587.0 & 347.4 & 56.5  \\
\hline
\multicolumn{4}{||c|}{ Case II		} \\
\hline
$e_t \equiv 0$ \hfill      & 1517.7  & 305.8 & 46.6  \\
\hline
\multicolumn{4}{||c|}{ Case III		} \\
\hline
$e_t \equiv 0$ \hfill      & 1487.8  & 295.9 & 46.3  \\
\hline
\multicolumn{4}{||c|}{ Case IV		} \\
\hline
$e_t=0.01$ \hfill	   & 1587.0 & 347.4 & 56.5  \\
$e_t=0.05$ \hfill	   & 1575.8 & 344.9 & 56.1  \\
$e_t=0.10$ \hfill	   & 1541.8 & 337.3 & 54.8  \\
\hline
\multicolumn{4}{||c|}{ Case V 	} \\
\hline
$e_t=0.01$ \hspace*{2.9cm} & 1829.5 & 497.8 & 91.9 \\
$e_t=0.05$ \hspace*{2.9cm} & 1817.3 & 494.7 & 91.4 \\
$e_t=0.10$ \hspace*{2.9cm} & 1779.2 & 485.1 & 89.6 \\
\hline
\multicolumn{4}{||c|}{ Case VI	} \\
\hline
$e_t\equiv 0$ \hfill       & 1830.5 & 497.4 & 91.9 \\
\hline
\end{tabular}
\end{table}

\section{Accurate GW phasing for eccentric binaries at 2.5PN order}
\label{sec: 2.5_GW_phasing}

In what follows, we briefly summarize a prescription to efficiently perform 2.5PN 
accurate GW phasing
for compact binaries of arbitrary mass ratio moving along inspiralling eccentric orbits
presented in Ref.~\cite{DGI}.
The temporal evolution of the dynamical variables, $r(t)$,
$\dot r(t)$, $\phi(t)$ and $\dot \phi(t)$, appearing in Eq.~(\ref{hx_ecc}) for
$h_{\times}|_Q(t)$,  is achieved by employing a version of the general Lagrange method
of variation of arbitrary constants. The idea is to split the compact binary dynamics,
defined by the relative acceleration $\mathcal A$, into two parts.
In general,    $\mathcal A$ consists of a conservative, and in our case integrable,  part
$\mathcal A_0$ and a reactive part $\mathcal A'$ that perturbes the conservative dynamics.
To perform efficient GW phasing, we employ a ``semi-analytic''
solution to the conservative dynamics $\mathcal A_0$. 
Thereafter, a solution to the PN accurate
dynamics  $\mathcal A = \mathcal A_0 + \mathcal A'$  is obtained by varying the constants
present in the generic solution to $\mathcal A_0$.

In the present paper, we restrict $\mc A_0$ to be 2PN accurate and  $\mc A'$ contains only
the dominant 2.5PN contributions, the so called Newtonian RR terms.   When
the conservative binary dynamics is 2PN accurate, it is convenient to express the relative
separation vector $\vek r$ (in a suitably defined `center-of-mass frame') as
\begin{equation}
\label{r_COM}
 \vek r = r\,\cos \phi \, \, {\vec {\bf i}} + r\, \sin \phi \, {\vec {\bf j}}\,,
\end{equation}
where ${\vec {\bf i}} = {\vec {\bf p}}$, 
~ ${\vec {\bf j}} = \cos i \, {\vec {\bf q}} + \sin i \, {\vec {\bf N}}$
[note that (${\vec {\bf p}}, {\vec {\bf q}}, {\vec {\bf N}}$) is the same orthonormal triad
employed to construct $h_{\times}|_Q$]. The 2PN
accurate expressions for the dynamical variables $r(t)$, $\phi(t)$ and their time derivatives
may be expressed parametrically as
\begin{subequations}
	\label{orb_elts}
\begin{align}
	r(t)		&=	r~(u(l), n, e_t) \\
	\dot r(t)		&=	\dot r~(u(l), n, e_t) \\
	\phi(t)		&=	\lambda (t, n, e_t)  + W(u(l), n, e_t) \\
	\dot \phi (t)	&=	\dot \phi~(u(l), n, e_t)
\end{align}
\end{subequations}
The explicit 2PN accurate expressions for the above quantities in harmonic gauge,
employed in this paper, can be extracted from Eqs.~(23), (24), (25) and (26) in
Ref.~\cite{KG06} and they are explicitely provided in 
appendix~\ref{phasing:Appendix:A}.
The basic angles $l$ and $\lambda$, symbolically given in Eq.~(\ref{phi_split})
[see Eqs.~(25) and (27) of Ref.~\cite{KG06} for the 2PN accurate explicit expressions], read
\begin{subequations}
 \begin{align}
	l	&=	n\,(t - t_0) + c_l \,, \\
	\lambda	&=	(1 + k) \, n \, (t - t_0) + c_{\lambda} \,.
 \end{align}
\end{subequations}
The explicit time evolutions for $r,  \dot r,  \phi$ and $\dot \phi$  are provided by the
following 2PN accurate Kepler equation,  in harmonic gauge,  which connects   $l$ and $u$
(see Eq. (27) in Ref.~\cite{KG06}):
\begin{align}
\label{2PN_KE}
l=&~u - e_t \sin u \nonumber \\
  &+\frac{ \xi^{4/3} }{8 \sqrt{1 - e_t^2} (1 - e_t \cos u)}
\biggl \{
(15 \eta - \eta^2 ) e_t \sin u \sqrt{1 - e_t^2} \nonumber \\
  &+12 ( 5 - 2 \eta ) (v-u) ( 1 - e_t \cos u )\biggr \}\,.
\end{align}
In the above description for the conservative 2PN accurate dynamics,
we have four constants
of integration, namely $n, e_t, c_l$ and $c_\lambda$. 
When the orbital dynamics is fully 2.5PN accurate, i.e., $\mc A~=~\mc A_0 + \mc A'$, then
the above four constants of integration become time dependent. The differential equations
governing the four constants of integration can be computed by demanding same functional
forms for $r, \dot r $ and $\dot \phi$ even when the binary dynamics is fully 2.5PN
accurate.
Further, following Ref.~\cite{DGI}, the temporal evolution			
in each $n, e_t, c_l$ and $c_\lambda$ can be treated to consist of a slow drift component
and a rapidly oscillating part. 
Symbolically, the above mentioned split in the four
variables reads
\begin{equation}
	\label{split_const_ev}
	c_\alpha(t)	=	\bar c_\alpha (t)	+ \tilde c_\alpha (t)	\,,
\end{equation}
where $\alpha$ denotes one of the four constants of 2PN accurate orbital dynamics, namely,
$n,e_t, c_l$ and $c_{\lambda}$.
To our desired PN order,
it was demonstrated in Ref.~\cite{DGI} that 
${d \bar c_l}/{d t} = {d \bar c_\lambda}/{d t}=0$ and 
${d \bar n}/{d t}$~ and ~${d \bar e_t}/{d t}$ are given by
Eqs.~(\ref{n_et_evol_correct}).
For the current (and preliminary) numerical investigations, we neglected small amplitude fast oscillations
$\tilde n, \tilde e_t, \tilde c_\lambda$ and $\tilde c_l$ 
(however, their explicit expressions in harmonic gauge may
be obtained from Eqs.~(36) in Ref.~\cite{KG06}).
To do GW phasing, we proceed as follows. Using 2PN accurate expressions for $r, \dot r,
\phi$ and $\dot \phi$ and 2PN accurate
Kepler equation, extractable from Eqs.~(23)-(27) in Ref.~\cite{KG06},
we compute $r(t), \dot r (t), \phi(t) $ and $\dot \phi(t)$ using 
$n$ and $e_t$ to represent our PN accurate eccentric orbit.
After that, we numerically impose reactive evolutions in $e_t(t)$ and $n(t)$, defined by
Eqs.~(\ref{n_et_evol_correct}),    on the 2PN accurate orbital dynamics and compute the
associated $h_{\times}|_{\rm Q} (t)$ using Eq.~(\ref{hx_ecc}).   This is how we compute
$\hxqt$ with PN accurate orbital evolution.
  
      From the above discussions, it is clear that an efficient implementation of GW phasing
also depends on an efficient (and accurate) way of solving the 2PN accurate Kepler Equation.
This is achieved by employing a slightly modified method of Mikkola to solve  the classical
Kepler Equation. This is detailed in the next subsection.

\subsection{ Mikkola's solution adapted to 2PN accurate Kepler Equation}

There exists a plethora of analytical and numerical solutions associated with the
celebrated classical KE, namely $l \equiv n (t-t_0) = u - e_t\, \sin u $
(see Ref.~\cite{KE_book}).
Arguably the most accurate and efficient numerical way to solve the classical KE is by
Seppo Mikkola \cite{SM}. Therefore, we adapt Mikkola's simple and robust method to
solve our 2PN accurate KE, given by Eq.~(\ref{2PN_KE}). Let us now 
we briefly discuss Mikkola's procedure. Recall that 
a numerical solution to the KE usually employs Newton's method which requires an initial
guess $u_{0}$ that depends on $l$ and $e_{t}$.  A number of iterations will be
required to obtain an approximate solution that has some desired accuracy.  The number of
iterations to reach this accuracy naturally depends on $u_{0}$, $e_t$ and $l$.

      In  Mikkola's method, $u_{0}$ is computed by introducing a new auxiliary variable
$s=\sin u/3$. The resulting equation for $s$ is Taylor expanded, forming a cubic polynomial
in $s$. The resulting approximate solution is empirically corrected to negate the largest error
occurring at $l=\pi$ leading to a $u$ value which has an accuracy of not less than
$10^{-3}$.  This accuracy is greatly enhanced by employing a fourth order version of Newton's
method (see Ref.~\cite{SM} for further details). This procedure provides $u$ as a function
of $l$ within an accuracy of $10^{-15}$ for all values of $e_t$ and $l$ for $e_t \in (0,1)$.
We note that the above method requires reduction of $l$ into the interval
$-\pi \le l \le \pi$ in order to make $s$ as small as possible. It is possible to employ
geometrical interpretations of $u$ and $l$, to map any $l$ in to the above interval, as
explained in Ref.~\cite{TG06}.
We note, while passing, that Mikkola's solution only demands the solution of a cubic
polynomial that involves no trigonometric functions and one-time evaluation of a couple
of simple trigonometric functions.

 To solve the fully 2PN accurate KE, we apply Mikkola's method at various PN orders
in a successive manner in the
following way. First we solve the 1PN accurate KE, $l=n(t-t_0)=u-e_t \sin u$, via Mikkola's
method, and obtain $u$ (note that due to the use of $e_t$ both 1PN and Newtonian  accurate
Kepler Equations have identical structure and hence solutions). This allows us to obtain
2PN corrections appearing on the right hand side of Eq.~(\ref{2PN_KE}) for $l$ in terms of
$(n,e_t, l)$ and let us denote it by $l_4$. We obtain the solution to 2PN accurate KE by
introducing $l' = l - l_4 = u-e_t \sin u$ ~and applying Mikkola's scheme to obtain $u(l')$.
This allows us to get accurate and efficient solution to 2PN accurate KE that provides $u$
in terms of $l, e_t, n, m $ and $\eta$.

  We are now in a position to explore data analysis implications of $h_{\times}|_{Q} (t) $
having 2.5PN accurate orbital motion.

\section{Nonoptimal Matched Filter Searches for GW signals having Some Residual 
Eccentricities}
\label{sec: NO_matched_filter}
 
In this section, we explore the implications for initial LIGO if
$h_{\times}\big|_{\rm Q}(t)$, given by Eq.~(\ref{hx_ecc}), evolving under 2.5PN
accurate orbital motion, represents a potential GW signal. This is done by revisiting the
analysis done in Ref.~\cite{MP00}.

   GW data analysts employ the technique of `matched filtering' to search for GWs from
compact binaries. This is an optimal technique provided one can construct search templates that
accurately model expected signals from GW sources,
especially in their phase evolution.
It is natural that these templates
depend on a number source parameters  and therefore one needs to construct a
`bank of templates' that densely covers the parameter space of a specific (and desired) GW
source.  The efficiency of a specific template bank in detecting the associated GW sources
is expressed by computing its Fitting Factor (FF), which is a measure of the `overlap'
between the expected GW signal and its approximate model, given by a template in the
template bank \cite{A95}. If a template bank can provide FFs $\sim 1$, one can be confident
about  the possible detection of the associated GW signals. Further, it was pointed out in
Ref.~\cite{A95} that the loss in event rate due to the deployment of inaccurate templates
is $\propto $ 1 - FF$^3$. And, the typical desirable 
FFs for initial LIGO are $\approx 0.97 $ \cite{BO96}.

  Following Refs.~\cite{A95,MP00}, we compute FFs in the following way. Let   
GW signals from compact binaries inspiralling along non-circular orbits 
be denoted by $s(t)$
and let  
the associated quasi-circular templates be
$ h(t;\vek{\lambda})$, where $\vek{\lambda}$ stands for template parameters. 
Further, let $ \tilde s(f)$ and $ \tilde h(f;\vek{\lambda})$ denote Fourier Transforms of $s(t)$
and $ h(t;\vek{\lambda})$, computed by employing  `realft' routine of \cite{NR}. With these inputs
and following Ref.~\cite{A95}, we define the ambiguity function $\cal A (\vek{\lambda})$ as
\be
{\cal A (\vek{\lambda})} = \frac{(s|h(\vek{\lambda}))}{\sqrt{(s|s)(h(\vek{\lambda})|h(\vek{\lambda}))}}
\,,
\ee
where the matched filter inner product $(a | b)$  by definition reads
\be
(a|b) = 2 \int_0^{\infty} \frac{\tilde a^*(f) \tilde b(f) + \tilde a(f) \tilde b^*(f)}{S_n(f)}\,df
\,.
\ee
In the above expression, $S_n(f)$ stands for the one sided noise power spectral density of the detector
and we have for 
the initial LIGO
\begin{subequations}
\label{spect_noise_densy}
\begin{align}
S_n(f)	=& S_0 [(4.49 x)^{-56}
		+0.16 x^{-4.52}
		+0.52 \nonumber \\
	 &	+0.32 x^2],	~~f \ge f_s \,, \\
	=&	\infty,		~~f <   f_s \,,
\end{align}
\end{subequations}
where $x=f/f_0$, with $f_0=150$Hz and $f_s=40$Hz. The constant $S_0$ never enters our computations
and hence its value is irrelevant for us.
When one employs quasi-circular templates, given by 
Eqs.~(\ref{hx_circ}) and (\ref{omega_evol}),
it is possible to separate $ \vek{\lambda}$ into $\vek{\lambda}=(t_0, \phi_0, \vek{\theta})$,
where $t_0$ and $\phi_0$ denote the time of arrival and the associated orbital phase of the
binary and $\vek{\theta}$ denotes the remaining parameters. Following Ref.~\cite{A95}, one can
easily maximize $\cal A$ over $t_0$ and $\phi_0$
while employing `realft' routine 
 to compute $ \tilde s(f)$ and $ \tilde h(f,\vek{\lambda})$.  
The FF is the maximum value of the ambiguity function, i.e, FF $\equiv ^{\rm max}_{~\vek{\lambda}} 
\cal A (\vek{\lambda}) $. For circular templates, defined by
Eqs.~(\ref{hx_circ}) and (\ref{omega_evol}), the only component of $\vek{\theta}$ is 
the chirp mass $\mathcal M $ and therefore
\be
FF = ^{\rm max}_{ t_0, \phi_0,\mathcal M} {\cal A}( \vek{\lambda})\,.
\ee
However, while employing the cases II, III and IV to model the circular templates,
it is clear that the above maximization has to be performed over $m$ and $\eta$.
For these cases, we invoked the mini-max overlap, detailed in Ref.~\cite{DIS98},
and the `amoeba' routine of Ref.~\cite{NR} to compute the associated FFs.
Below we describe how we tackled the analysis performed in Ref.~\cite{MP00}.

\subsection{ Brief summary of how we revisited the analysis of Ref.~\cite{MP00}}

 Let us first briefly describe how GW signals were actually constructed in Ref.~\cite{MP00}.
They employed Newtonian accurate conservative orbital evolution and imposed
on  it  the  adiabatic evolution of orbital elements arising from Einsteinian quadrupole
formulae for the energy and angular momentum fluxes.
The actual construction of GW signals from compact binaries in inspiralling eccentric
orbits was achieved by employing semi-latus rectum  $p$ and eccentricity $e$ to describe
the Newtonian accurate (conservative) orbital motion.  The effects of radiation reaction
were  incorporated by employing coupled differential equations for ${d p}/{dt} $ and
${de}/{dt}$, computed using energy and angular momentum balances.
We note that in Ref.~\cite{MP00} the construction of the GW signal, $s(t)= h_{\times}\big|_{\rm Q} (t)$,
required solving numerically three coupled differential equations, namely, those for
$ {d \phi}/{dt}, {d p}/{dt} $ and ${de}/{dt}$
(see Eqs.~(6), (8) and (9) in Ref.~\cite{MP00}).
Further, Ref.~\cite{MP00} computed frequency domain versions of Newtonian accurate
quasi-circular 
templates, using the stationary phase approximation
[see their Eqs.~(13) and (14) and (15)].

  To repeat the analysis of Ref.~\cite{MP00}, we constructed GW signals by
employing Eqs.~(\ref{hx_ecc}) for 
$h_{\times}|_{Q}(t) $, Eqs.(\ref{orbital_elmts_NEWT}) for the
purely Newtonian accurate conservative orbital motion and Eqs.~(\ref{n_et_evol_correct})
to impose the effects of radiation reaction on the orbital motion.
Therefore,  our prescription leads to solving numerically coupled differential equations
for  $n$ and $e_t$ along with the classical Kepler Equation.
Our quasi-circular templates  are given by Eqs.~(\ref{hx_circ}) and 
(\ref{omega_evol}) and we employed the `realft'
routine to compute $ \tilde s(f)$ and $ \tilde h(f,\vek{\lambda})$. 
For a given initial eccentricity, 
initial value for
mean motion $n = {2\, \pi}/{T} $ is chosen following 
Ref.~\cite{TG06},
which provided a detailed spectral analysis for GWs from eccentric binaries. This allowed
us to choose, for Newtonian binaries of all masses, $n_i = 125.7$Hz when $e_i=0.01$, the
initial value for $e_t$, and $ n_i = 83.8$Hz for $e_i = 0.1$.
Note that in Ref.~\cite{MP00}, initial orbital frequency was always chosen to be
$ 40/3 = 13.3$Hz, where $40$Hz being the lower frequency cut-off for the initial LIGO.
This is mainly due to the assumption of Ref.~\cite{MP00} that GWs from eccentric binaries 
can be decomposed into components that oscillate at once, twice and thrice the orbital 
(radial) frequency. 
We let the final value for $n$ be $ (6^{3/2}\, G\, m /c^3)^{-1} $
as compact binaries that we considered here were circularized when $n$ reached the
above value and because $n = \omega$ at Newtonian order.
Further, in most FF computations, the search templates were chosen to evolve in a GW
frequency window having 40Hz and $(6^{3/2}\,\pi \, G m/c^3)^{-1}$ as their lower and upper
limits. We noted that our results were somewhat insensitive to minor changes in $n_f$.
We are aware that Ref.~\cite{MP00} employed $h_{+}$ for their FF computations. However,
we choose $h_{\times}$ mainly because we can factor out $-2\,G\,m\,C/(c^2 R')$ from our
GW signals and templates.

  Employing the above mentioned procedure and restricting the initial values of $e_t \leq 0.1$,
we were able to reproduce most of the entries in Tables II and III of Ref.~\cite{MP00}.
This gave us a lot of confidence in our routines to implement the FFs. In the next subsection,
we detail the FF computations that employ Eq.~(\ref{hx_ecc}) having
2.5PN accurate orbital evolution to model GWs from eccentric binaries, while 
employing the cases I, II, III and VI as the search templates 
to probe the performances of these circular templates.


\subsection{Fitting Factor computations involving the cases I, II, III, V and VI
}
\label{subsec: NO filt _ FFs for GW}

 In this section, first, we present the results from our FF computations that employ 
Eq.~(\ref{hx_ecc}) having 2.5PN accurate orbital evolution, detailed 
in the Section~\ref{sec: 2.5_GW_phasing},
to model, somewhat realistically,  GWs from inspiralling eccentric binaries. 
And, we employ as search templates the 
three types of circular waveforms, given by Eqs.~(\ref{hx_circ}) and ~(\ref{omega_evol}),
Eqs.~(\ref{hx_circ}) and (\ref{Eq.3}) and  Eqs.~(\ref{hx_circ}) and (\ref{Eq.4}) [and these are
the above detailed cases I, II and III].
In other words, we are exploring if the adiabatic, complete adiabatic and 
gauge-dependent complete non-adiabatic circular templates, having reactive evolution 
restricted to the dominant qudrupolar order will be efficient to capture our fiducial eccentric  
signal, detailed in Section~\ref{sec: 2.5_GW_phasing}.
We are aware that comparing the cases V and I may be objectionable due to the use of
Newtonian accurate quantities in the derivation of the right hand side of Eq.~(\ref{Eq.2b}).
However, and as mentioned earlier,
 we pursued such a comparison due to the following two reasons.
First, it is interesting to note, from a strict PN point of view, that Eqs.~(\ref{omega_evol}) 
also provide 2.5PN order GW phase evolution. Secondly, we want to explore 
whether the various types
of circular templates, detailed in Ref.~\cite{AIRS}, can capture our somewhat realistic eccentric 
$h_{\times}(t)$. We would like to emphasize that a similar objection is \emph{not} 
applicable while employing the cases II and III to model circular templates. 
The FFs, relevant for the initial-LIGO, for the cases I, II and III while 
employing Eq.~(\ref{hx_ecc}) having 2.5PN accurate orbital phase evolution
to model the fiducial eccentric GW signals  
are listed in Table~\ref{Table:2_FF_MCH}.
\begin{table} [!ht]
\caption
{\label{Table:2_FF_MCH}
The FFs, relevant for the initial-LIGO, involving the adiabatic, complete adiabatic 
and gauge-dependent complete non-adiabatic approximants, namely the cases I, II and III,
as search templates and where the reactive order is at the dominant quadrupolar level.
The fiducial eccentric signals are constructed using Eq.~(\ref{hx_ecc}) where 
the orbital dynamics is fully 2.5PN accurate [the case V]. 
The templates that provide the listed FFs are characterized by 
$\mathcal M_t$, $m_t$ and $\eta_t$. 
We do not employ the third harmonic while modeling our eccentric GW signals.
}
\begin{tabular}{||l r|r|r|r|}
\hline
 $m_1/ M_{\odot} : m_2/ M_{\odot}$ &		& $1.4 : 1.4$ & $1.4 : 10$ & $10 : 10$	\\
\hline
\multicolumn{5}{||c|}{ Case~V Vs Case~I           } \\
\hline
$e_t=0.01$	&		\hspace*{1.5cm}	FF	& 0.533  & 0.509 & 0.655	\\
		&${\mathcal M_{t}}/{\mathcal M_s}$	& 0.910  & 0.802 & 0.745	\\
\hline
$e_t=0.05$	&		\hspace*{1.5cm} FF	& 0.532  & 0.507 & 0.654	\\
		&${\mathcal M_{t}}/{\mathcal M_s}$	& 0.910  & 0.802 & 0.744	\\
\hline
$e_t=0.10$	&		\hspace*{1.5cm} FF	& 0.522  & 0.507 & 0.659	\\
		&${\mathcal M_{t}}/{\mathcal M_s}$	& 0.912  & 0.803 & 0.746	\\
\hline
\multicolumn{5}{||c|}{ Case~V Vs Case~II           } \\
\hline
$e_t=0.01$	&		\hspace*{1.5cm}	FF	& 0.430  & 0.504 & 0.755	\\
		&${ m_t}/{m_s}$				& 0.870  & 0.622 & 0.614	\\ 
		&${ \eta_t}/{\eta_s}$			& 0.992	 & 1.366 & 0.987	\\ 
\hline
$e_t=0.05$	&		\hspace*{1.5cm}	FF	& 0.428	 & 0.500 & 0.744	\\ 
		&${ m_t}/{m_s}$				& 0.877	 & 0.655 & 0.659	\\ 
		&${ \eta_t}/{\eta_s}$			& 0.980	 & 1.250 & 0.863	\\ 
\hline
$e_t=0.10$	&		\hspace*{1.5cm}	FF	& 0.417	 & 0.493 & 0.732	\\ 
		&${ m_t}/{m_s}$				& 0.900	 & 0.689 & 0.623	\\ 
		&${ \eta_t}/{\eta_s}$			& 0.939	 & 1.151 & 0.934	\\ 
\hline
\multicolumn{5}{||c|}{ Case~V Vs Case~III           } \\
\hline
$e_t=0.01$	&		\hspace*{1.5cm}	FF	& 0.424	 & 0.500 & 0.772	\\ 
		&${ m_t}/{m_s}$				& 0.852	 & 0.580 & 0.601	\\ 
		&${ \eta_t}/{\eta_s}$			& 1.000	 & 1.492 & 0.997	\\ 
\hline

$e_t=0.05$	&		\hspace*{1.5cm}	FF	& 0.420	 & 0.494 & 0.762	\\ 
		&${ m_t}/{m_s}$				& 0.886	 & 0.678 & 0.662	\\ 
		&${ \eta_t}/{\eta_s}$			& 0.936	 & 1.139 & 0.836	\\ 
\hline
$e_t=0.10$	&		\hspace*{1.5cm}	FF	& 0.411	 & 0.487 & 0.752	\\ 
		&${ m_t}/{m_s}$				& 0.889	 & 0.737 & 0.603	\\ 
		&${ \eta_t}/{\eta_s}$			& 0.935	 & 0.984 & 1.000	\\ 
\hline
\end{tabular}
\end{table}

Significantly lower FFs, listed in Table~\ref{Table:2_FF_MCH}, 
clearly indicate that GW templates, defined 
by Eqs.~(\ref{hx_circ}), (\ref{omega_evol}),
(\ref{Eq.3}) and (\ref{Eq.4})
are not efficient in capturing our somewhat 
realistic GW signals.
For the sake of completeness, we note that 
FF computations that used GW signals having lower values of initial $e_t$ also
gave us similar lower numbers [we note that it is rather difficult (numerically) to use 
values below
$10^{-3}$ for initial eccentricities in our GW signals construction].  
However, low FFs are consistent with the arguments, based on the number of 
accumulated GW cycles, presented in Section~\ref{sec:MP00}.
  
 Let us now explain the way we constructed the inspiral templates associated with the case VI.
The new circular inspiral template family is obtained by taking 
the circular limit, defined by $ e_t \rightarrow 0 $,
of Eqs.~(6), (26), (48d), (51), (52), (53) and (63) in Ref.~\cite{DGI}.
Using these expressions in the limit $ e_t \rightarrow 0 $, 
we constructed a new type of GW templates 
for compact binaries inspiralling along PN accurate quasi-circular 
orbits. The relevant expressions are
\begin{align}
\label{quasi_c_hx}
h_{\times} ( \phi, n) |_R &=  -4 C \frac{(G m \eta) }{c^2 R'} 
\biggl ( \frac{G m n}{c^3}  \biggr )^{2/3}
 \sin 2\phi \,,
\end{align}
where $ \phi(t) $ and $n(t) $ are governed by 
\begin{subequations}
\label{quasi_c_dt_phi}
\begin{align}
\label{Eq.20a}
\frac{d \phi}{dt} &= n \biggl \{ 1 + 3\,\xi^{2/3} + \left( {\frac {39}{2}} - 7\, \eta
\right ) \, {\xi}^{4/3} \biggr \}\,,
\\
\label{Eq.20b}
\frac{d n}{dt} &= \frac{96}{5} \, \eta \, \left ( \frac{ G m n}{c^3} \right )^{5/3} n^2 \,.
\end{align}
\end{subequations}
In our expression for $h_{\times} ( \phi, n) |_R$ we have neglected PN corrections to
$ {G\, m}/{r} $ and $\dot \phi$ in the  $ {e_t  \rightarrow 0} $ limit.
This is justified as we can
treat these corrections forming PN contributions to the amplitude of $h_{\times}$ (this is
indeed the way PN corrections to $h_{\times}$ and $h_+$ were computed in Ref.~\cite{BIWW}).
We detail below various properities of this new approximant.

  It is interesting to note that, in the PN approximation
and while dealing with the restricted PN waveforms, Eqs.~(\ref{quasi_c_hx}) and 
(\ref{quasi_c_dt_phi})
are simply another representation of Eqs.~(\ref{hx_circ}) and (\ref{omega_evol}).
In other words, both Eqs.~(\ref{hx_circ}) and (\ref{omega_evol})  and
Eqs.~(\ref{quasi_c_hx}) and (\ref{quasi_c_dt_phi}) can be used to describe GWs from 
quasi-circular
compact binaries having 2.5PN accurate non-stationary orbital phase 
evolution. The difference is that the use of $d \phi/dt \equiv \omega$ leads to a PN independent 
stationary phase evolution and while in terms of $n$, it is evidently 
PN dependent. 
Further, following \cite{DS88},
one can also demonstrate that $n$ is indeed a gauge invariant PN accurate quantity like
$\omega$. Therefore, based on purely theoretical arguments, it is not possible to
argue against the use of $n$ to construct quasi-circular 
PN accurate GW search templates. In principle, 
one can employ other gauge invariant quantities like the orbital energy or angular momentum to
construct GW search templates. 
We have used $n$ in Eqs.~(\ref{quasi_c_hx}) and (\ref{quasi_c_dt_phi})
as 2.5PN accurate GW phasing for eccentric binaries, available in Ref.~\cite{DGI}, employed
$n$.
And, what are the most appropriate variables to do perform accurate GW phasing
for eccentric binaries will be reported elsewhere.

 A rather confusing consequence of expressing $ d \phi/dt = \omega = n(1 +k)$
in terms of $n$ for the quasi-circular inspiral is the following. The PN accurate conservative
angular motion becomes $ \phi - \phi_{0} = n \left ( 1 + k \right ) (t - t_0)$.
Therefore, deviations from the Newtonian accurate angular motion can be attributed
the rate of periastron advance. This is despite of the fact that for circular orbits,
it is rather difficult to visualize the periastron advance. However, if one employs, for example,
the orbital energy to prescribe the orbit, it is no more possible to make the above
statement as both $n$ and $k$ are PN accurate functions of the orbital energy.
We note that similar arguments and expressions were employed in Ref.~\cite{BDI95} during
the PN accurate GW phasing of quasi-circular inspiral [see Eqs.~(4.20)--(4.24) in Ref.~\cite{BDI95}
and the appendix~\ref{phasing:Appendix:A}]. 

 Finally, we would like to point out that Eqs.~(\ref{quasi_c_dt_phi}) also represent 
an adiabatic inspiral. This is because the balance arguments, similar to the ones employed 
in the derivation of $d \omega/dt$ and $dx/dt$, are also required to compute $dn/dt$.
In contrast, in the complete non-adiabatic approximant, one does not require to invoke 
any balance arguments.
Further, it is possible to provide the following physical explanation for the new template family.
One may state that Eqs.~(\ref{quasi_c_hx}) and (\ref{quasi_c_dt_phi}) model 
GWs from a compact binary in a 2PN accurate circular orbit, defined by Eq.~(\ref{Eq.20a}), perturbed by 
the dominant order radiation reaction, given by Eq.~(\ref{Eq.20b}).
We would like to recall that in Ref.~\cite{DGI}, the phasing for eccentric binaries were performed
by perturbing compact binaries in PN accurate eccentric orbits by 
appropriate PN accurate reactive dynamics for eccentric binaries. 

 Therefore,  we repeated our above mentioned FF computations
using $h (t, \vek{\lambda}) $, given by Eqs.~(\ref{quasi_c_hx}) and
(\ref{quasi_c_dt_phi}) and we again employed the `realft' routine to compute
the associated  $\tilde s(f)$ and $\tilde h(f, \vek{\lambda})$. Our results for three
canonical compact binaries having  three initial eccentricities, $ e_t = 0.01, 0.05$ and
$0.1$ are listed in Table.~\ref{Table:3_FF_m_eta}.
For mildly eccentric binaries, having initial $e_t \sim 0.1$, there are two ways of 
computing FFs. In the first case, we neglected the existence of the third harmonic which 
allow us to use Eq.~(\ref{omega_n_2PN}) with $ \omega = 40\, \pi$Hz while computing $n_i$.
Following Ref.~\cite{TG06}, we may include the third harmonic by adding $n/2$ to the right hand side 
of Eq.~(\ref{omega_n_2PN}) and equating it to $40\, \pi$Hz while numerically evaluating $n_i$.
The associated FFs are also given in Table~\ref{Table:3_FF_m_eta}.

\begin{table}
\caption{
\label{Table:3_FF_m_eta}
The initial-LIGO FFs that uses  Eqs.~(\ref{quasi_c_hx}) and (\ref{quasi_c_dt_phi})
to model circular GW templates and Eqs.~(\ref{hx_ecc}) having 2.5PN accurate orbital motion to
represent fiducial eccentric GW signals.
The symbols $m_t$ and $\eta_t$, as expected, denote the total mass and the symmetric mass
ratio of the desired template.
The initial and final values of $\omega$ and $n$ are chosen to have $f_{GW}$ varying
between 40 Hz and $(6^{3/2}\, \pi \, G m/c^3)^{-1}$. For our $m_1=m_2=1.4 M_\odot$ binary, the
orbital evolution is terminated when the relevant harmonic reaches 1000Hz: the photon
shot noise limit. When the initial $e_t=0.1$, the FFs are computed by including (and neglecting) the
third harmonic.
We include the effect of the third harmonic by using, for example, $n_i=78.76$Hz for
GWs signals having $m=11.4$ and $\eta=0.108$.
}
\begin{tabular}{||l r|r|r|r|}
\hline
$m_1/ M_{\odot} : m_2/ M_{\odot}$ &	& $1.4 : 1.4$ 	& $1.4 : 10.0$ 		& $10.0 : 10.0$ \\
\hline
$e_t=0.01$& \hspace*{0.8cm}	FF		& 0.999 & 0.999 & 0.999	\\
$	$&			$m_t $		& 2.813 & 11.42 & 20.16	\\
$	$&			$\eta_t$	& 0.248 & 0.107 & 0.247	\\
\hline
$e_t=0.05$&			FF		& 0.966 & 0.993 & 0.992	\\
$	 $&			$m_t $		& 2.937 & 11.72 & 20.40	\\
$	 $&			$\eta_t$	& 0.233 & 0.104 & 0.244	\\
\hline
\multicolumn{5}{||c|}{Without the third harmonic}\\
\hline
$e_t=0.10$&			FF		& 0.894	& 0.965 & 0.970	\\
$	 $&			$m_t$		& 3.133	& 12.71 & 20.66	\\
$	 $&			$\eta_t$	& 0.212	& 0.094 & 0.244	\\
\hline
\multicolumn{5}{||c|}{Including the third harmonic}\\
\hline
$e_t=0.10$&			FF		& 0.888 & 0.983 & 0.984	\\
$	 $&			$m_t $		& 2.833 & 11.69 & 20.51	\\
$	 $&			$\eta_t$	& 0.246 & 0.102 & 0.244	\\
\hline
\end{tabular}
\end{table}

\begin{figure}[ht]

\caption{
\label{FIG:ambiguity_eta_mt}
The overlap for a binary system with
$m_1=1.4 M_{\odot}$, $m_2=10.0 M_{\odot}$ and initial
$e_t=0.1$ in the $m_t$~-~$\eta_t$ space.
It is visually rather difficult to pinpoint a global maximum. However, for a large number
of $(m_t, \eta_t)$ values, the overlap is more than 0.92.
}
\centering
\includegraphics[height=8cm, angle=0]{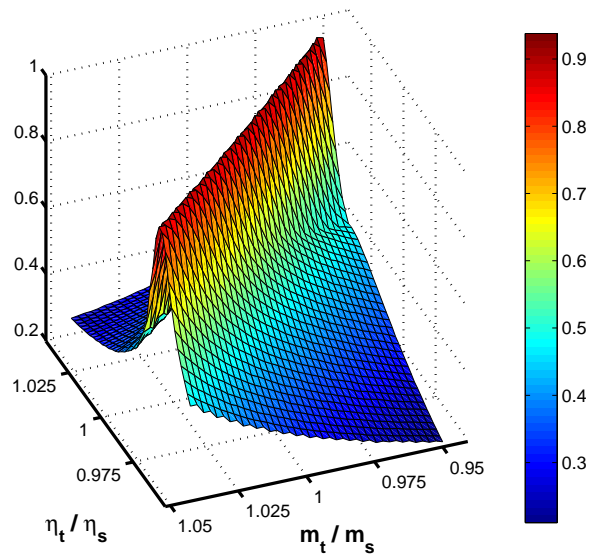}
\end{figure}

The numbers presented in the Table~\ref{Table:3_FF_m_eta},
in comparison with the Table.~\ref{Table:2_FF_MCH},
clearly indicate that the quasi-circular GW templates, given by
Eqs.~(\ref{quasi_c_hx}) and (\ref{quasi_c_dt_phi}), are highly 
efficient in capturing GWs from compact binaries having some residual orbital
eccentricity.
The higher FFs for massive compact binaries should be due to the fact that they emit 
stronger GWs and have smaller time window to dephase.
Finally, in Fig.~\ref{FIG:ambiguity_eta_mt}, we plot $\mathcal{O} ( \vek{\theta}) $, the so-called overlap,
obtained by maximizing the ambiguity function over the kinematical variables
$t_0$ and $\phi_0$.

 To make sure that Eqs.~(\ref{quasi_c_hx}) and (\ref{quasi_c_dt_phi})
 representing our quasi-circular GW templates, can lead to higher FFs,
we performed the following test.
It is obvious that ${dn}/{dt}$, given in Eqs.~(\ref{quasi_c_dt_phi}), 
can be written in terms of $\cal M $ as
\bea
\frac{d n}{dt} &= \frac{96}{5} \, \left ( \frac{ G\, {\cal M}\,n}{c^3} \right )^{5/3} n^2 \,.
\eea
Therefore, we constructed a set of quasi-circular GW templates that used the same $m$ and $\eta$ 
as our PN eccentric GW signal to define $ {d \phi}/{dt}$, given in  Eq.~(\ref{quasi_c_dt_phi}).
However, we used above equation for $dn/dt$ and treated 
$\cal M$ as an independent parameter and computed FFs while
maximizing over $\cal M$ [and this is much simpler procedure than maximizing over 
$m$ and $\eta$]. The resulting FFs turned out to be very close
to the entries listed in Table~\ref{Table:3_FF_m_eta}.

 We conclude the section by providing following brief summary of our analysis.
When one restricts radiation reaction to the dominant order and wants to model 
GWs from compact binaries in eccentric orbits, it is appropriate and desirable to employ Ref.~\cite{DGI}.
We have demonstrated that certain
circular templates, given by Eqs.~(\ref{quasi_c_hx}) and (\ref{quasi_c_dt_phi}),
representing GWs from compact binaries inspiralling under quadrupolar radiation reaction
along 2PN accurate circular orbits, are efficient in capturing our fiducial GW signals from eccentric 
binaries. The crucial difference between the various circular templates, considered as the cases I, II and III,and the case VI is that the 
Eqs.~(\ref{quasi_c_hx}) and (\ref{quasi_c_dt_phi}), representing the case VI, treat the conservative 
and reactive orbital phase evolution in a gauge-invariant manner with equal emphasis.



\section{Concluding Discussions}
\label{sec: conclusion}

We explored the drops in the signal-to-noise-ratios, relevant for the initial LIGO, while
non-optimally searching for GWs from compact binaries inspiralling under quadrupolar
radiation reaction along PN accurate
eccentric orbits with various types of circular templates.
The fiducial GW signals are somewhat realistically modeled 
with the help of Ref.~\cite{DGI},
that provided $h_{\times,+}|_{Q}(t)$ from inspiralling eccentric binaries having 2.5PN
accurate adiabatic phase evolution.
We demonstrated that search templates obtained by perturbing compact binaries in 2PN 
accurate circular orbits with dominant order reactive dynamics are highly efficient in 
capturing  our GW signals even from very mildly eccentric compact binaries. 
However, the search templates arising from the adiabatic, complete adiabatic and 
complete non-adiabatic approximants were found to be rather inefficient 
in capturing the fiducial GW signals having even tiny  residual orbital eccentricities. 
In our view, the present analysis indicates that it is desirable to treat, in an equal footing,
 both the conservative and reactive contributions to the orbital phase evolution, while constructing
the search templates for GWs from astrophysical compact binaries.

  We are aware that the various kinds of PN accurate search templates for the quasi-circular inspiral,
available in LAL \cite{LAL}, employ directly or indirectly 
Eq.~(\ref{Eq.3a}) and different prescriptions for the PN-accurate reactive evolution of $x(t)$.
Therefore, it will be rather important to extend our present analysis to investigate
if various quasi-circular search templates, employed by the GW data analysts, can capture
GW signals from compact binaries inspiralling along eccentric orbits
and, say, having 3PN accurate conservative and 2PN accurate reactive dynamics.
For such an analysis, the construction of PN accurate GW signals from inspiralling
eccentric binaries should be influenced by Refs.~\cite{DGI,KG06} and this is currently
under active investigation.
Further, while providing more accurate prescriptions
to compute GW signals from eccentric binaries, 
it is imperative to include currently neglected
contributions due to $\tilde c_{\alpha}$, appearing at 2.5 and 3.5PN orders.
We also feel that such post Newtonian accurate versions of the present analysis should be repeated
for VIRGO, Advanced LIGO and LISA.
It is also desirable to explore the effects of the spins on the present FF computations.
This requires construction of GW signals from spinning compact binaries inspiralling along PN
accurate eccentric orbits building on a preliminary investigation presented in Ref.~\cite{KG05}.

\begin{acknowledgments}

We are very grateful to 
Gerhard~Sch\"afer for detailed 
discussions and encouragements.
We thank anonymous referee and Eric Poisson for their opinions.
This work is supported in part by the grants from 
the DFG (Deutsche Forschungsgemeinschaft) through SFB/TR7
``Gravitationswellenastronomie'' and
the DLR (Deutsches Zentrum f\"ur Luft- und Raumfahrt).

\end{acknowledgments}

\appendix

\section{2PN accurate orbital dynamics for eccentric binaries}
\label{phasing:Appendix:A}

 In this appendix, we provide explicit 2PN accurate parametric expressions for 
$r, \dot r, \phi $ and $\dot \phi$ in harmonic gauge, required to construct
$h_{\times,+}|_{Q}(t)$ from inspiralling eccentric binaries having 2.5PN
accurate orbital motion. These parametric expressions are extracted from 
Eqs.~(23)-(26) in Ref.~\cite{KG06}.
The radial motion, parametrically
defined by $r(l,n,e_t)$ and $\dot{r}(l,n,e_t)$,
reads 
\begin{widetext}
\begin{subequations}
\label{phasing_eq:23}
\begin{align}
\label{phasing_eq:23a}
r & 
= r_{\rm N}
+ r_{\rm 1PN}
+ r_{\rm 2PN}
\,,
\\
\dot{r} & 
= \dot{r}_{\rm N}
+ \dot{r}_{\rm 1PN}
+ \dot{r}_{\rm 2PN}\,,
\end{align}
\end{subequations}
where 
\begin{subequations}
\begin{align}
r_{\rm N} & =
\left( \frac{G M}{n^2} \right)^{1/3} ( 1 - e_t \cos u )
\,,
\\
r_{\rm 1PN} & =
r_{\rm N} \times
\frac{ \xi ^{2/3} }{ 6 ( 1 - e_t \cos u ) }
[ - 18 + 2 \eta - ( 6 - 7 \eta ) e_t \cos u ]
\,,
\\
r_{\rm 2PN} & = r_{\rm N} \times
\frac{ \xi^{4/3} }{ 72 (1 - e_t^2 ) (1 - e_t \cos u) }
\Bigl \{
- 72 (4 - 7 \eta )
+ \left[
72 + 30 \eta + 8 \eta^2 - ( 72 - 231 \eta + 35 \eta ^2 ) e_t \cos u
\right] (1 - e_t^2 )
\nonumber
\\
& \quad
- 36 ( 5 - 2 \eta ) (2 + e_t \cos u ) \sqrt{1 - e_t^2}
\Bigr \}
\,,
\\
\dot{r}_{\rm N} & =
\frac{ (G M n)^{1/3} }{ (1 - e_t \cos u) } e_t \sin u
\,,
\\
\dot{r}_{\rm 1PN} & =
\dot{r}_{\rm N} \times
\frac{ \xi^{2/3} }{6} ( 6 - 7\eta )
\,,
\\
\dot{r}_{\rm 2PN} & =
\dot{r}_{\rm N} \times
\frac{ \xi^{4/3} }{ 72 ( 1 - e_t \cos u)^3 }
\biggl [
- 468 - 15 \eta + 35 \eta^2
+ ( 135 \eta - 9 \eta^2 ) e_t^2
+ ( 324 + 342 \eta - 96 \eta^2 ) e_t \cos u
+ ( 216 - 693 \eta
\nonumber
\\
& \quad
+ 105 \eta^2 ) ( e_t \cos u )^2
- ( 72 - 231 \eta + 35 \eta^2 ) ( e_t \cos u )^3
+ \frac{ 36 }{ \sqrt{1 - e_t^2} }
( 1 - e_t \cos u )^2 ( 4 - e_t \cos u ) ( 5 - 2 \eta )
\biggr ]
\,,
\end{align}
\end{subequations}
\end{widetext}
The angular motion, describable using PN accurate parametric expressions for
$\phi$ and $\dot{\phi}$,
is given by
\begin{widetext}
\begin{subequations}
\label{phasing_eq:25}
\begin{align}
\label{phasing_eq:25a}
\phi(\lambda,l) & = \lambda + W(l)
\,,
\\
\label{phasing_eq:25b}
\lambda & = ( 1 + k ) l
\,,
\\
\label{phasing_eq:25c}
W(l) &
= W_{\rm N}
+ W_{\rm 1PN}
+ W_{\rm 2PN}
\,,
\\
\dot{\phi} &
= \dot{\phi}_{\rm N}
+ \dot{\phi}_{\rm 1PN}
+ \dot{\phi}_{\rm 2PN}\,,
\end{align}
\end{subequations}
\end{widetext}

 The above mentioned various PN quantities read
\begin{widetext}
\begin{subequations}
\begin{align}
k &
= \frac{ 3 \xi ^{2/3} }{ 1 - e_t^2 }
+ \frac{ \xi ^{4/3} }{ 4 ( 1 - e_t^2 )^2 }
\left[ 78  - 28 \eta + ( 51 - 26 \eta ) e_t^2 \right]
\,,
\\
W_{\rm N} &
= v - u + e_t \sin u
\,,
\\
W_{\rm 1PN} &
= \frac{ 3 \xi^{2/3} }{ 1 - e_t^2 }  ( v - u + e_t \sin u )
\,,
\\
W_{\rm 2PN} &
= \frac{ \xi ^{4/3} }{ 32 ( 1 - e_t^2 )^2 ( 1 - e_t \cos u )^3 }
\bigg (
8
\left[
78 - 28 \eta
+ ( 51 - 26 \eta) e_t^2
- 6 ( 5 - 2 \eta ) ( 1 - e_t^2 )^{3/2}
\right] ( v - u ) ( 1 - e_t \cos u )^3
\,,
\\
\dot{\phi}_{\rm N} & =
\frac{ n \sqrt{1 - e_t^2} }{ (1 - e_t \cos u )^2 }
\,,
\\
\dot{\phi}_{\rm 1PN} & =
\dot{\phi}_{\rm N} \times
\frac{ \xi^{2/3} }{ (1 - e_t^2) ( 1 - e_t \cos u) }
\left[ 3  - ( 4 - \eta ) e_t^2  + ( 1 - \eta ) e_t \cos u \right]
\,,
\\
\dot{\phi}_{\rm 2PN} & =
\dot{\phi}_{\rm N} \times
\frac{ \xi^{4/3} }{ 12 (1 - e_t^2)^2 ( 1 - e_t \cos u)^3 }
\biggl \{
144
- 48 \eta
- ( 162 + 68 \eta - 2 \eta^2 ) e_t^2
+ ( 60 + 26 \eta - 20 \eta^2 ) e_t^4
+ ( 18 \eta + 12 \eta^2 ) e_t^6
\nonumber
\\
& \quad
+ \left[
- 216
+ 125 \eta
+ \eta^2
+ ( 102 + 188 \eta + 16 \eta^2 ) e_t^2
- ( 12 + 97 \eta - \eta^2 ) e_t^4
\right] e_t \cos u
+ \left[
108
- 97 \eta
- 5 \eta^2
\right.
\nonumber
\\
& \quad
\left.
+ ( 66 - 136 \eta + 4 \eta^2 ) e_t^2
- ( 48 - 17 \eta + 17 \eta^2 ) e_t^4
\right] ( e_t \cos u )^2
+ \left[
- 36
+ 2 \eta
- 8 \eta^2
- ( 6 - 70 \eta - 14 \eta^2 ) e_t^2
\right] ( e_t \cos u )^3
\nonumber
\\
& \quad
+ 18
( 1 - e_t \cos u )^2
( 1 - 2 e_t^2 + e_t \cos u )
( 5 - 2 \eta )
\sqrt{1 - e_t^2}
\biggr \}
\,.
\end{align}
\end{subequations}
\end{widetext}


In the circular limit, it is not that difficult to conclude that the
2PN accurate expression for $\dot \phi$ becomes
\begin{widetext}
\begin{align}
\label{omgn}
\dot \phi &=
n \biggl \{ 1 + 3\,{\xi}^{2/3}
+ \left( {
\frac {39}{2}}-7\,\eta \right) {\xi}^{4/3}
\biggr \}\,.
\end{align}
\end{widetext}
The above equation, in terms of $k$, is simply $\dot \phi = n\,\left (1 +k \right )$. 
Therefore, while using $n$ and in the limit
$e_t \rightarrow 0$, deviations from the Newtonian accurate orbital motion may be 
explained in terms of $k$. However, this explanation is representation dependent as 
shown below.
Using PN accurate expressions, given in Ref.~\cite{MGS},
one can easily represent in the circular limit $\dot \phi $ 
using the conserved orbital energy and it reads
\begin{widetext}
\begin{align}
\label{omgE}
\dot \phi &= 
\frac{ \zeta^{3/2}}{G\,m} \biggl \{
1 + \frac{1}{8} \left[ {9}+\eta \right] \frac{\zeta}{c^2}
+ \biggl [
{\frac {891}{128}}
-{\frac {201 }{64}}\,\eta
+{\frac {11}{128}}\,{\eta}^{2}
\biggr ] \frac{ {\zeta}^{2}}{c^4}
\biggr \}
\,,
\end{align}
\end{widetext}
where $\zeta = -2\, E/\eta\,m $ and $E$ being the conserved center-of-mass orbital energy.
While employing $\zeta$, it is obvious that deviations from 
Newtonian accurate phase evolution can not be
explained purely in terms of $k$. 

A comparison of Eqs.~(\ref{omgn}) and (\ref{omgE}) leads to the following observations.
Using the usual definition $\dot \phi = \omega = 2\,\pi/T$ and Eq.~(\ref{omgE}), one may
define a PN accurate period $T$. However, at PN orders, this is not given by $T_r=2\,\pi/n$, 
the radial period of Keplerian type parameterization
and the period relevant to construct GW signals from inspiralling eccentric binaries.
The above defined $T$ is related to $n$ by
$ T = \frac{2\, \pi}{ n\,\left(1 +k \right)}$.
Further, observe that there are various PN accurate ways of defining $\omega$ 
and they, at a given GW phasing order, lead to various quasi-circular GW templates.


\begin{thebibliography}{999}


\bibitem{Abbott:2007wu}
B.~Abbott {\it et al.} [LIGO Scientific Collaboration],
[arXiv:gr-qc/07040943].

\bibitem{Hild:2006bk}
S.~Hild  [LIGO Scientific Collaboration],
 Class.\ Quant.\ Grav.\  {\bf 23}, S643 (2006).

\bibitem{Acernese2006}
F.~Acernese {\it et al.},
Class.\ Quant.\ Grav.\  {\bf 23}, S635 (2006).

\bibitem{BDI}
T.~Damour, P.~Jaranowski, and G.~Sch\"afer, Phys.\ Lett.\ B
\textbf{513}, 147 (2001);
L.~Blanchet, G.~Faye, B.~R.~Iyer, and B.~Joguet,
Phys.\ Rev.\ D \textbf{65}, 061501(R) (2002); \textbf{71}, 129903(E) (2005);
L.~Blanchet, T.~Damour, G.~Esposito-Far\`ese, and B.~R.~Iyer,
Phys.\ Rev.\ Lett. \textbf{93}, 091101 (2004);
K.~G.~Arun, L.~Blanchet, B.~R.~Iyer, and M.~S.~S.~Qusailah,
Class.\ Quant.\ Grav. \textbf{21}, 3771 (2004); \textbf{22}, 3115(E) (2005)
and references therein.

\bibitem{P64}
P.C.~Peters, Phys.\ Rev.\textbf{136}, B1224 (1964).

\bibitem{MP00}
K.~Martel and E.~Poisson,
Phys.\ Rev.\  D {\bf 60}, 124008 (1999)

\bibitem{e_scenarios}
H.~K.~Chaurasia and M.~Bailes,
Astrophys.\ J. \textbf{632}, 1054 (2005); K.~L.~Page \textit{et al.},
Astrophys.\ J. \textbf{637}, L13 (2006); 
J.~Grindlay, S.~P.~Zwart, and S.~McMillan,
Nature (London) \textbf{2}, 116 (2006).

\bibitem{LAL}
\emph{\bibinfo{title}{Lsc algorithm library {(LAL)}}},
\url{http://www.lsc-group.phys.uwm.edu/lal};
T.~Damour, B.~R.~Iyer and B.~S.~Sathyaprakash,
Phys.\ Rev.\  D {\bf 63} (2001) 044023
[Erratum-ibid.\  D {\bf 72} (2005) 029902]
[arXiv:gr-qc/0010009].


\bibitem{AIRS}
P.~Ajith, B.~R.~Iyer, C.~A.~K.~Robinson and B.~S.~Sathyaprakash,
  Phys.\ Rev.\  D {\bf 71}, 044029 (2005)
  [Erratum-ibid.\  D {\bf 72}, 049902 (2005)]
  [arXiv:gr-qc/0412033].

\bibitem{DGI}
T.~Damour, A.~Gopakumar, and B.~R.~Iyer,
Phys.\ Rev.\ D \textbf{70}, 064028 (2004).

\bibitem{DD}
T.~Damour and N.~Deruelle,
Ann. Inst. Henri Poincar\'e Phys. Th\'eor. \textbf{44}, 263 (1986).

\bibitem{DS88}
T.~Damour and G.~Sch\"afer,
Nuovo Cimento Soc. Ital. Fis., B \textbf{101}, 127 (1988).

\bibitem{WS}
G.~Sch\"afer and N.~Wex,
Phys. Lett. \textbf{174 A}, 196, (1993);
erratum \textbf{177}, 461.

\bibitem{MGS}
R.-M.~Memmesheimer, A.~Gopakumar, and G.~Sch\"afer,
Phys.\ Rev.\ D \textbf{70}, 104011 (2004).

\bibitem{KG06}
C.~K\"onigsd\"orffer and A.~Gopakumar,
Phys.\ Rev.\  D {\bf 73}, 124012 (2006).


\bibitem{TG06}
M.~Tessmer and A.~Gopakumar,
Mon.\ Not.\ Roy.\ Astron.\ Soc.\  {\bf 374}, 721 (2007)
[arXiv:gr-qc/0610139].

\bibitem{KE_book}
P. Colwell, {\it Solving Kepler's Equation Over Three Centuries}, 
Willman-Bell, Inc. (1993).


\bibitem{SM}
S.~Mikkola,
{\it Celestial Mechanics}, {\bf 40}, 329 - 334, (1987).

\bibitem{A95}
T.~A.~Apostolatos,
Phys.\ Rev.\  D {\bf 52}, 605 (1995).

\bibitem{BO96}
  B.~J.~Owen,
  Phys.\ Rev.\  D {\bf 53}, 6749 (1996)
  [arXiv:gr-qc/9511032].


\bibitem{NR}
W.~H.~Press, W.~T.~Vetterling,S.~A.~Teukolsky and B.~P.~Flannery,
{\it Numerical Recipes in C++: the art of scientific computing},
Cambridge University Press, (2002).


\bibitem{DIS98}
T.~Damour, B.~R.~Iyer and B.~S.~Sathyaprakash,
Phys.\ Rev.\  D {\bf 57}, 885 (1998)
[arXiv:gr-qc/9708034].


\bibitem{BIWW}
L.~Blanchet, B.~R.~Iyer, C.~M.~Will, and A.~G.~Wiseman,
Class.\ Quant. Grav.\ \textbf{13}, 575 (1996).


\bibitem{BDI95}
L.~Blanchet, T.~Damour, and B.~R.~Iyer,
Phys.\ Rev.\ D \textbf{51}, 5360 (1995).

\bibitem{KG05}
C.~K\"onigsd\"orffer and A.~Gopakumar,
Phys.\ Rev.\ D \textbf{71}, 024039 (2005).


















\end{thebibliography}
\end{document}